\documentclass[%
12pts,
english,
showpacs,
reprint,
epsf,
superscriptaddress,
aps,
prb,
floatfix,
]{revtex4-2}

\usepackage[utf8]{inputenc}
\usepackage{hyperref}           
\usepackage{marvosym}
\usepackage{amssymb}   
\usepackage{bbm}
\usepackage{babel}
\usepackage{bm}    
\usepackage{color}
\usepackage{natbib}
\usepackage{graphicx,subfigure} 
\usepackage{epstopdf}
\usepackage{float}
\usepackage{soul}
\usepackage{color,graphicx}
\graphicspath{ {images/} }
\usepackage{dcolumn}    
\usepackage{wrapfig}
\usepackage{mathtools}
\usepackage{xspace}
\usepackage{amsmath}
\usepackage{refstyle}   
\usepackage{tikz}  
\usetikzlibrary{arrows,shapes,trees,positioning}  
\usepackage{comment} 
\usepackage{siunitx}

\usepackage{appendix}
\usepackage{hyperref}            
\usepackage{textcomp}
\usepackage{booktabs}
\usepackage[mathlines]{lineno}

\usepackage{stmaryrd}
\usepackage[normalem]{ulem}
\usepackage[ignoreunlbld,norefs,nocites]{refcheck}
\usepackage{ulem}
\usepackage{multirow}
\hypersetup{
	colorlinks=true,
	linkcolor=blue,
	citecolor=blue,
	urlcolor=blue,
}
\DeclareUnicodeCharacter{2212}{-}

\definecolor{darkgreen}{rgb}{0, 0.8, 0}

\newcommand{\FFGT}{Fe\texorpdfstring{$_4$}{4}GeTe\texorpdfstring{$_2$}{2}\xspace}

\newcommand{\FthGT}{Fe$_{3}$GeTe$_{2}$\xspace}

\newcommand{\Aog}{A$_{1g}$\xspace}
\newcommand{\Eg}{E$_{g}$\xspace}

\newcommand{\TC}{$T_\mathrm{C}$\xspace}
\newcommand{\TSR}{$T_\mathrm{SR}$\xspace}

\begin{document}


 \title{Spin Reorientation Driven Renormalization of Spin–Phonon Coupling in \FFGT}


\author{Riju Pal}
\email{rijupal07@gmail.com}
\affiliation{Department of Condensed Matter and Materials Physics, S. N. Bose National Centre for Basic Sciences, Block JD, Sector III, Salt Lake, Kolkata 700106, India}

\author{Md. Nur Hasan}
\affiliation{Department of Physics and Astronomy, Uppsala University, Box 516, SE-751 20 Uppsala, Sweden}

\author{Chumki Nayak}
\affiliation{Department of Physical Sciences, Bose Institute, 93/1, Acharya Prafulla Chandra Road, Kolkata 700009, India}

\author{Mrinal Deka}
\affiliation{Department of Physics, Indian Institute of Technology Madras, Chennai 600036, India}

\author{Nastaran Salehi}
\affiliation{Department of Physics and Astronomy, Uppsala University, Box 516, SE-751 20 Uppsala, Sweden}

\author{Manuel Pereiro}
\affiliation{Department of Physics and Astronomy, Uppsala University, Box 516, SE-751 20 Uppsala, Sweden}

\author{Suchanda Mondal}
\affiliation{Saha Institute of Nuclear Physics, 1/AF Bidhannagar, Calcutta 700064, India}

\author{Abhishek Misra}
\affiliation{Department of Physics, Indian Institute of Technology Madras, Chennai 600036, India}

\author{Achintya Singha}
\affiliation{Department of Physical Sciences, Bose Institute, 93/1, Acharya Prafulla Chandra Road, Kolkata 700009, India}

\author{Prabhat Mandal}
\email{prabhatmandalphysics@gmail.com}
\affiliation{Department of Condensed Matter and Materials Physics, S. N. Bose National Centre for Basic Sciences, Block JD, Sector III, Salt Lake, Kolkata, 700106, India}

\author{Debjani Karmakar}
\email{karmakar.debjani@gmail.com}
\affiliation{Department of Physics and Astronomy, Uppsala University, Box 516, SE-751 20 Uppsala, Sweden}
\affiliation{Homi Bhabha National Institute, Anushaktinagar, Mumbai, 400094, India}
\affiliation{Technical Physics Division, Bhabha Atomic Research Centre, Mumbai 400085, India}

\author{Atindra Nath Pal}
\email{atin@bose.res.in}
\affiliation{Department of Condensed Matter and Materials Physics, S. N. Bose National Centre for Basic Sciences, Block JD, Sector III, Salt Lake, Kolkata, 700106, India}

\date{\today}       


\begin{abstract}
Quasi-2D van der Waals ferromagnet Fe$_4$GeTe$_2$, featuring the simultaneous presence of high Curie temperature (\TC $\sim 270$ K) and a spin-reorientation transition at \TSR $\sim 110$ K, is a rare system where strong interplay of spin dynamics, lattice vibrations, and electronic structure leads to a wide range of interesting phenomena. Here, we investigate the lattice response of exfoliated Fe$_4$GeTe$_2$ nanoflakes using temperature-dependent Raman spectroscopy. Polarization-resolved measurements reveal that, while one Raman mode exhibits a purely out-of-plane character, the rest display mixed symmetry, reflecting interlayer vibrational nonuniformity and symmetry-driven mode degeneracies. Below \TC, phonons harden, and the linewidth narrows, consistent with reduced anharmonicity, while across the spin reorientation transition at \TSR\ they display anomalous softening, linewidth broadening, and a peak in lifetime, which are signatures of strengthened spin-phonon coupling. Complementary DFT+DMFT calculations and atomistic spin dynamical simulations reveal temperature-dependent spin excitations whose energies overlap with the Raman-active phonons, providing a natural route for the observed magnon-phonon interaction. Together, these insights establish Fe$_4$GeTe$_2$ as a versatile platform for exploring intertwined spin, lattice, and electronic degrees of freedom, with relevance for dynamic spintronic and magneto-optic functionalities near technologically meaningful temperatures.

\end{abstract}

\maketitle

\section*{Introduction}

Coupling between spin and lattice degrees of freedom, known as spin--phonon coupling, plays a pivotal role in shaping the collective behavior of correlated magnetic materials by governing how magnetic order and lattice vibrations mutually respond to each other \cite{Hu2023, Sun2021}. In systems with multiple magnetic sublattices, the interplay between site-dependent magnetic moments and the crystal lattice can lead to nonlinear (higher-order) spin--lattice interactions~\cite{crosstalk_Natcomm, May2019c}, which stabilize long-range order or drive competing magnetic and structural phases~\cite{Son2019,saha_spin-lattice-magnetodielctric}. When dimensionality is reduced, as in layered or two-dimensional (2D) magnets, quantum confinement and modified electronic correlations amplify these effects, giving rise to enhanced spin--phonon interactions and emergent magnetostructural phenomena~\cite{Raman_review1, Huang2020, Yin2020a, Hu2023, Subhadeep-FePS3, Subhadeep_Spin-phonon-heterostructure, pradeep_CGT, saha-electron-phonon-nips3, saha-nips3-shortrangecorrelation, samanta2024spin}. Magnetocrystalline anisotropy, especially uniaxial, together with spin–orbit coupling, dictates the preferred orientation of magnetic moments and their interaction with lattice vibrations. In addition, the intrinsic symmetry breaking in 2D systems further strengthens spin–lattice coupling, leading to phonon softening and other vibrational anomalies. For example, in CrI$_3$, spin-orbit coupling stabilizes a specific magnetization direction \cite{Sivadas2018, Huang2020raman}, while in Cr$_2$Ge$_2$Te$_6$ it affects both thermal conductivity and magnetization by altering the phonon spectrum~\cite{Tian2016}. Such magnetoelastic interactions can also drive dynamic magnetostructural transitions \cite{FePS31}, inducing exotic states such as skyrmions and ferroelectricity \cite{Huang2020}. 

\begin{figure*}[t!]
	\centering
	\includegraphics[width=1.0\textwidth]{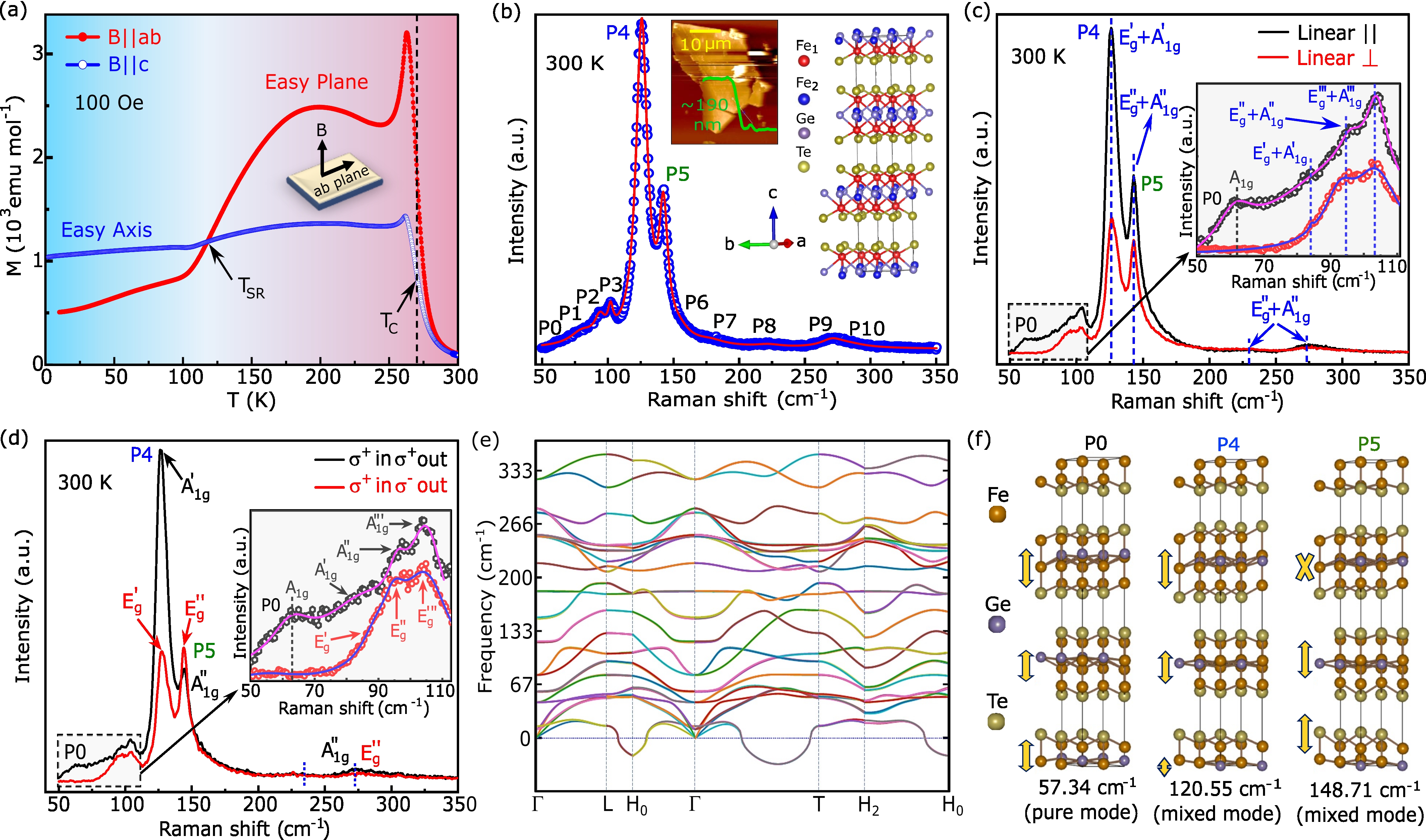}
	\caption{\textbf{Characterization of Raman modes of \FFGT at 300~K.} \textbf{(a)} Temperature-dependent magnetization data of \FFGT (at 100 Oe) showing ferromagnetic transition at \TC $\sim$ 270 K and spin-reorientation transition at \TSR. \textbf{(b)} Raman spectrum (blue circles) of an \FFGT\ flake measured with $\lambda_{\mathrm{ex}}=532$ nm, fitted using eleven Raman modes (P0–P10). The fitted curve is shown in red. Upper left inset: AFM height profile and image of the flake showing corresponding thickness of $\sim$190 nm. Right inset: Crystal structure of \FFGT, consisting of seven-atom-thick monolayers stacked in an ABC sequence, forming the rhombohedral structure (space group R$\bar{3}$m, No. 166). \textbf{(c)} Linear polarization–resolved Raman spectra (inset: magnified view of P0–P3). \textbf{(d)} Helicity-resolved Raman spectra at 300 K (inset: magnified view of P0–P3). \textbf{(e)} Phonon dispersion of \FFGT, calculated using Phonopy-VASP. \textbf{(f)} Calculated phonon frequencies and representative vibrational patterns for selected modes of \FFGT\ obtained from Phonopy–VASP, highlighting mode-specific symmetry and atomic displacements.}
	\label{fig:Fig1}
\end{figure*}

In metallic ferromagnets Fe$_n$GeTe$_2$ ($n$ = 3, 4, 5) with high transition temperatures, the presence of conduction electrons complicates the interplay between lattice and spin dynamics~\cite{May2019c, Kong2021, Pal_2024_aip, Pal2024, Seo2020b, PalESR2024, Bera2022f3gt}. Interaction between charge carriers and phonons not only alters the electronic structure but also mediates magnetoelastic effects that significantly impact magnetic ordering \cite{Du2019}. Among these compounds, Fe$_4$GeTe$_2$ (\TC $\sim 270$ K) stands out for its spin-reorientation transition at \TSR $\sim 110$ K, where the magnetic easy axis reorients from out-of-plane to in-plane with increasing temperature (see $M-T$ curve in Fig.~\ref{fig:Fig1} (a))~\cite{Pal2024, Seo2020b, PalESR2024, Mondal2021, Bera2023}. This transition is accompanied by unconventional magnetic and magnetotransport behavior, including maxima in both the negative magnetoresistance and anomalous Hall response, along with a sign reversal of the ordinary Hall coefficient~\cite{Bera2023, Pal2024, mondal2022anomalous}. Recent studies incorporating the impacts of dynamical electronic correlation and field-induced symmetry breaking have indicated a complex magnetic phase diagram of the system~\cite{PhysRevB.111.134449}. Electron spin resonance (ESR) studies reveal a temperature-driven evolution of spin-wave modes closely tied to electronic structure~\cite{PalESR2024}. Understanding magnon–phonon and phonon–phonon interactions is thus key to unraveling the temperature evolution of these properties. 

Raman spectroscopy is a sensitive tool to probe the phonon dynamics and their coupling to magnetic excitations~\cite{Wei2025, Du2025, saha-prb-spin-phonon, saha-nips3-shortrangecorrelation, zhang2025, Subhadeep-FePS3, Kim2019a}. Polarization-resolved Raman spectra of exfoliated \FFGT nanoflakes reveal conventional phonon hardening and linewidth narrowing upon cooling through the ferromagnetic phase, consistent with reduced anharmonic scattering. As the temperature approaches \TSR, however, several Raman-active modes exhibit pronounced phonon softening and linewidth broadening, indicating enhanced spin-phonon coupling. First-principles calculations, combining phonon analysis with DFT+DMFT-based spin dynamical simulations, determine the symmetry of the Raman modes and a temperature evolution of spin wave excitations whose energies overlap with these phonons, supporting the observed anomalies at \TSR.

\section*{Symmetry-Resolved Lattice Dynamics in \FFGT}

\FFGT\ crystallizes in a rhombohedral structure (space group $R\bar{3}m$, No.166) with trigonal symmetry and $D_{3d}$ point group~\cite{Seo2020b, Mondal2021, Pal2024} (see \emph{inset}, Fig.\ref{fig:Fig1}(b)). Group theory predicts Raman-active modes of \Aog\ and doubly degenerate \Eg\ symmetries~(see Section S1 and Table T1). Their polarization dependence of intensity ($I$) is governed by the Raman tensors $\bm{R}$, $I \propto |\bm{p_i} \cdot \bm{R} \cdot \bm{p_s}|^2$, where $\bm{p_i}$ and $\bm{p_s}$ are the incident and scattered light polarization vectors \cite{Loudon2001}. 

\begin{figure*}[t!]
	\centering
	\includegraphics[width=0.95\textwidth]{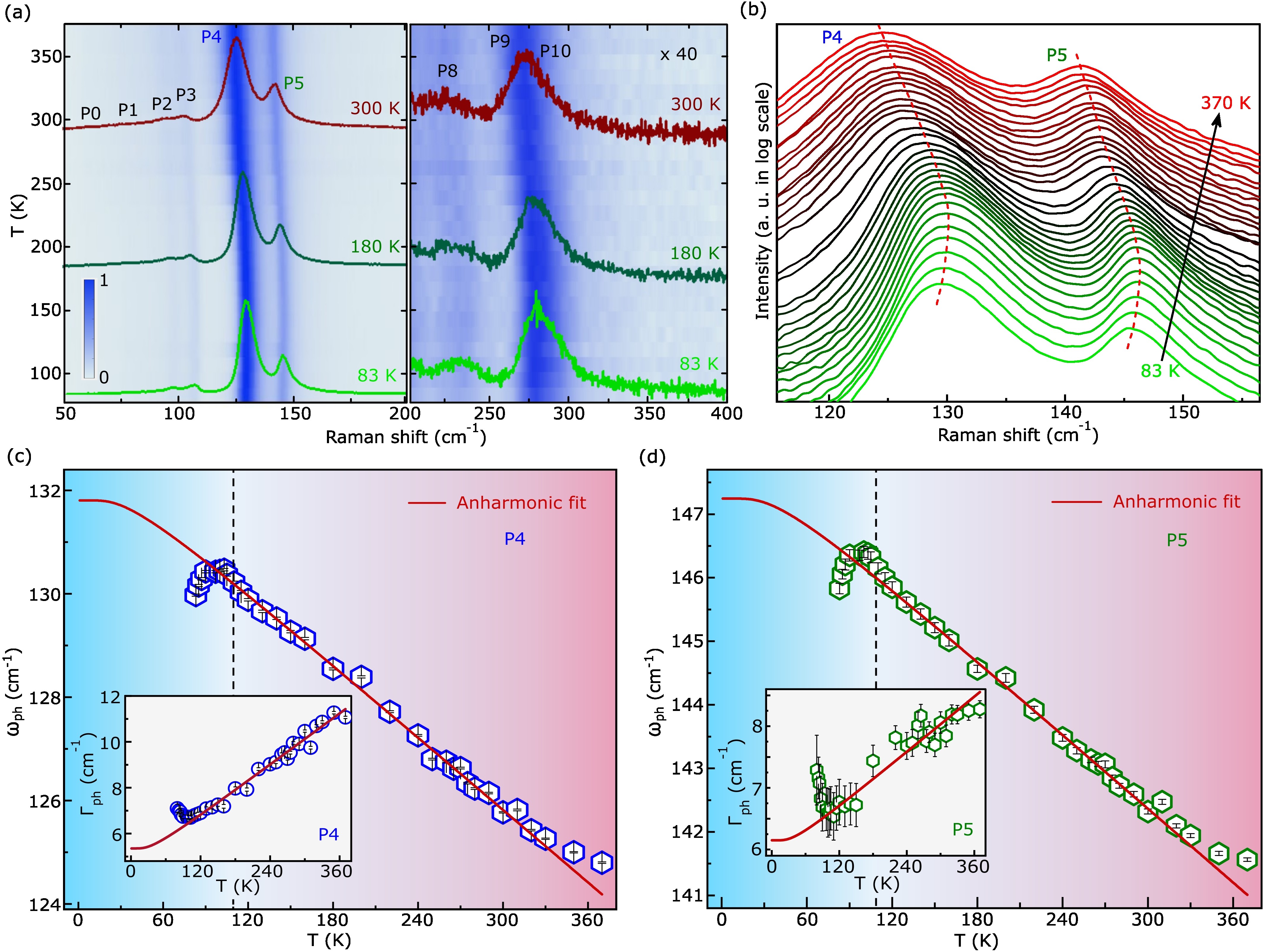}
	\caption{\textbf{Temperature dependent Raman response of \FFGT.} \textbf{(a)} Two-dimensional density map of Raman intensity as a function of temperature, along with representative spectra at 83, 180, and 300 K. Blue denotes maximum spectral intensity, while white corresponds to zero intensity. Each spectrum is individually normalized within the 200–400 cm$^{-1}$ range. \textbf{(b)} Semi-logarithmic plot of temperature-dependent Raman spectra highlighting the systematic shifts of the dominant modes (P4 and P5); the red dashed line serves as a guide to the eye.  \textbf{(c,d)} Temperature evolution of the phonon frequencies $\omega_{\mathrm{ph}}$ for modes P4 and P5. Both modes exhibit clear deviations from the standard anharmonic behavior (red fits; see text). Insets: Temperature-dependent phonon linewidths for the same modes, fitted to the anharmonic model in the 120–300 K range.}
	\label{fig:Fig2}
\end{figure*}

\begin{table*}[htb]
    \textbf{Table 1: Phonon energies (in cm$^{-1}$) of \FFGT at 300 K obtained from experimental data and theoretical calculations}
    \vspace{0.1 cm}
    \label{Table1}
	\centering
	\renewcommand{\arraystretch}{2.0}
    \resizebox{1\textwidth}{!}{
	\begin{tabular}{@{}cccccccccccccc@{}}
		\toprule
        \hline
		\textbf{Phonons: (cm$^{-1}$)} & \textbf{P0} & \textbf{P1} & \textbf{P2} & \textbf{P3} & \textbf{P4} & \textbf{P5} & \textbf{P6} & \textbf{P7} & \textbf{P8} & \textbf{P9} & \textbf{P10} \\
		\hline
  
        \textbf{Experiment:} & 61 $\pm$ 2.6 & 80.4 $\pm$ 0.51 & 94 $\pm$ 0.2 & 102 $\pm$ 0.1 & 125.8 $\pm$ 0.02 & 142.3 $\pm$ 0.03 & 154.8 $\pm$ 0.7 & 179 $\pm$ 1.7 & 222 $\pm$ 2.9 & 270.8 $\pm$ 0.9 & 283 $\pm$ 2.4 \\

       \textbf{Theory}: & 57.34 & 78.13 & 78.47 & 119.3 & 120.55 & 148.71 & 150.48 & 183.3 & 218.3 & 280.74 & 286.3 \\
        \hline
        \midrule
	\end{tabular}
    }
\end{table*}

We examined the lattice dynamics using Raman spectroscopy on thin layers of \FFGT\ exfoliated onto Si/SiO$_2$ substrates. To prevent oxidation, the samples were transferred to a high-vacuum environment immediately after exfoliation for measurements. Raman data were collected from flakes with thickness down to 8~nm ($\sim$8 layers) using various excitation wavelengths (see Section S2 and Table~T2 for details). However, the main manuscript focuses on a $\sim$190~nm thick flake (R190) that exhibits strong and well-defined peaks (see AFM image in Fig. \ref{fig:Fig1}(b), upper inset). Thinner flakes show broader peaks due to surface oxidation. Nonetheless, application of higher laser power could remove the oxidized layer and restore the characteristic Raman features of \FFGT (see Section S3 and S4) \cite{laser_ok1, laser_ok2, laser_ok4, laser_ok3}. Measurements were conducted in backscattering geometry using linearly polarized light, with the incident wave vector aligned along the crystallographic c-axis. We used two cryostats: a liquid nitrogen flow cryostat covering 83–370~K, and a pulse-cooled dry cryostat operating between 5–300~K, though the latter had lower spectral resolution (see Section S5).

\begin{figure*}[t!]
	\centering
	\includegraphics[width=0.99\textwidth]{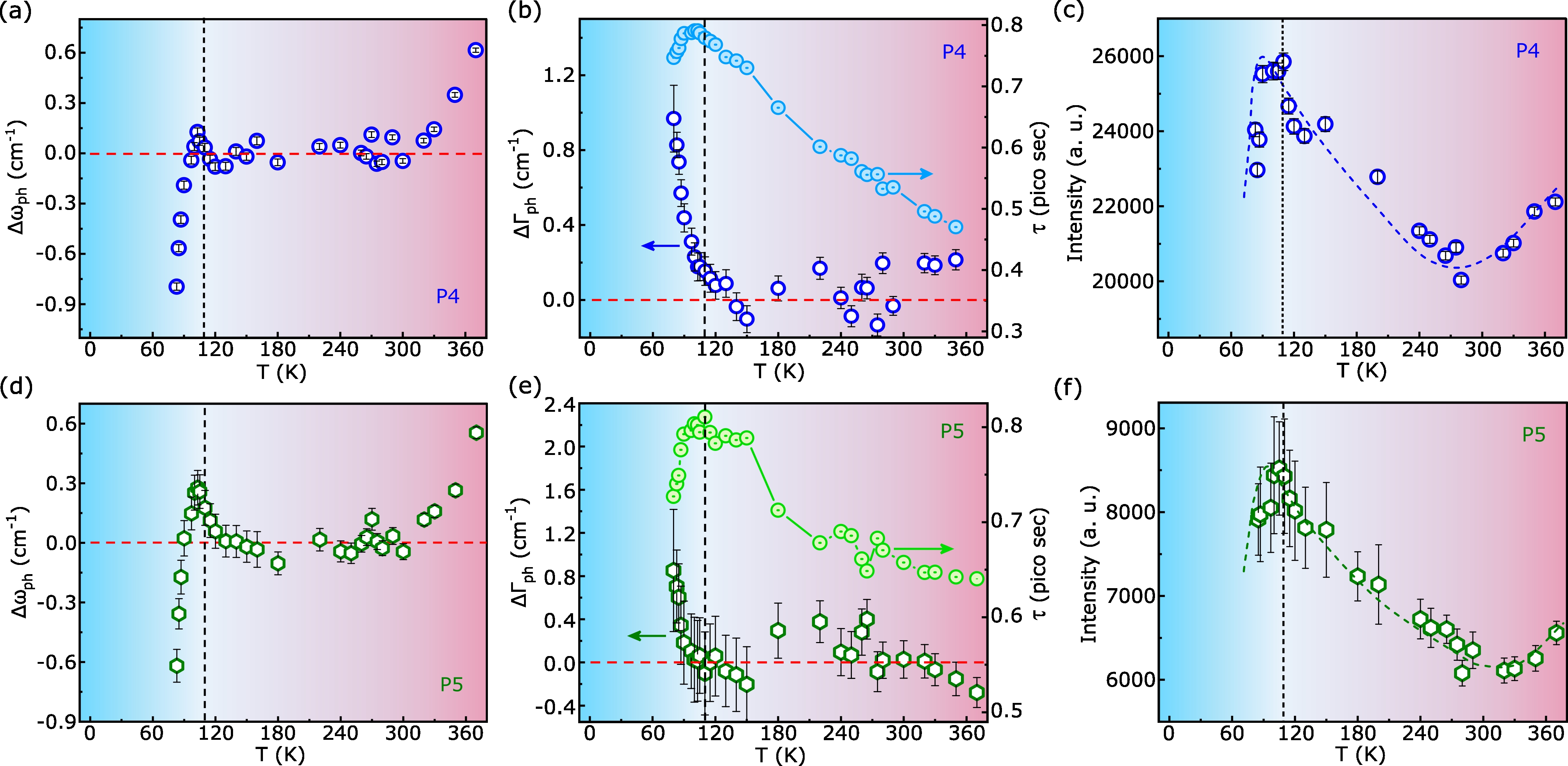}
    \caption{\textbf{Temperature-dependent deviations of phonon properties.} Temperature evolution of the phonon frequency deviations $\Delta \omega_{\mathrm{ph}}(T) = \omega \mathrm(T) - \omega _\mathrm{anh}(T)$ \textbf{(a,d)} and linewidth deviations $\Delta\Gamma_{\mathrm{ph}}(T) = \Gamma \mathrm(T) - \Gamma _\mathrm{anh}(T)$ \textbf{(b,e)} (left axes) from the standard anharmonic model for the Raman modes P4 and P5. Both modes exhibit pronounced deviations below \TSR\ and minor deviations above $\sim$\TC. The right axes in \textbf{(b,e)} show the temperature dependence of the phonon relaxation time $\tau$ for P4 and P5, displaying marked changes around \TSR. The color gradient highlights distinct temperature regions across \TSR, consistent with the $M$–$T$ behavior. The black dotted line marks \TSR. \textbf{(c,f)} Temperature-dependent changes in the mode intensities of P4 and P5. Dashed lines serve as guides to the eye.}
		\label{fig:Fig3}
\end{figure*}

Fig.~\ref{fig:Fig1}(b) shows the room-temperature Raman spectrum of R190, measured under 2.33 eV laser excitation. No significant features appear above 350 cm$^{-1}$ (see Section~S6).  Deconvolution of the spectrum using multiple Lorentzian fits reveals eleven Raman modes (P0–P10), labeled in Fig. \ref{fig:Fig1}(b) and detailed in Section S6.2. The most intense modes, P2 ($\sim$ 94 cm$^{-1}$), P3 (102 cm$^{-1}$), P4 (125.8 cm$^{-1}$), and P5 (142.3 cm$^{-1}$), are accompanied by seven weaker features: P0 (61), P1 (80.4), P6 (154.8), P7 (179), P8 (222.4), P9 (270.8), and P10 (283.3 cm$^{-1}$) (see Table 1). For further analysis, we focus on modes P2–P5 due to their strong sensitivity to temperature and polarization, as well as their reliable spectral quality.

To determine the symmetry of the observed phonon modes, Raman measurements were performed for configurations with linear and circular polarization. In case of linearly polarized spectra (Fig.~\ref{fig:Fig1}(c)), P0 appears exclusively in the co-polarized geometry, whereas all other prominent modes (P1–P5 and P8–P10) are observed in both co- and cross-polarized configurations, though the intensity is markedly higher in the co-polarized case. Helicity-resolved spectra (Fig.~\ref{fig:Fig1}(d)) exhibit a similar trend: P0 is present only in the co-circular channel, while the remaining modes appear in both co- and cross-circular geometries. Taken together, the linear and circular polarization unambiguously assign P0 to the A$_{1g}$ symmetry, whereas the modes P1–P5 and P8–P10 contain mixed contributions from A$_{1g}$ and E$_g$ symmetries.

For a comprehensive understanding of the lattice dynamics of \FFGT, we have carried out a systematic theoretical investigation of the phonon dispersions and the behavior of different phonon/vibrational modes over 10-320 K (Fig.~S25). The phonon dispersion data, as presented in Fig. \ref{fig:Fig1}(e), are calculated along the high-symmetry path $\Gamma$-L-$H_0$-$\Gamma$-T-$H_2$-$H_0$ of the Brillouin zone. To compare with the experimental data, we concentrated on the phonon modes at the zone center ($\Gamma$-point), and the closest theoretical value to the experimental Raman-active modes is presented in Table~1. The theoretically obtained phonon frequencies: 57.34, 78.13, 78.47, 119.3, 120.55, 148.71, 150.48, 183.3, 218.3, 280.74, and 286.3 $\text{cm}^{-1}$, correspond closely to the experimentally observed Raman peaks P0–P10 (see Fig.~\ref{fig:Fig1}(f) and Fig.~S26). These calculations capture the essential vibrational behavior of \FFGT\ and align well with the measured phonon energies, apart from a few modes that show noticeable deviations.

The nature of the atomic vibrations associated with the P0 to P10 phonon modes in \FFGT was analyzed computationally to elucidate the differences between the pure and mixed vibrational characters. Phonon dispersions were computed using two distinct structural models: (i) the experimental lattice structure without any ionic relaxation, and (ii) a fully relaxed structure. For the most intense experimental peaks (P2–P5), the calculated phonon frequencies of the relaxed structure match well with the P2 and P3 modes, but show significant deviations for P4 and P5. In contrast, the unrelaxed structure reproduces the experimental P4 and P5 frequencies more accurately, although the corresponding P2 and P3 modes appear shifted relative to experiment. Since P4 and P5 are the most intense experimental peaks, we concentrate on the unrelaxed computational data (Fig. \ref{fig:Fig1}(e)) for further analysis. To differentiate the characteristics of the pure and mixed vibrational modes (Table 1) in \FFGT, we have analyzed the nature of their atomic vibrations for phonon mode frequencies P0, P4, and P5, as depicted in Fig. \ref{fig:Fig1}(f) and in Supplementary Movie 1. The polarization-resolved Raman analysis identifies P0 as an out-of-plane vibrational mode. Our calculations predict a mode at 57.34 $\text{cm}^{-1}$, which closely aligns with the experimental observation for P0. The rest of the higher-frequency modes (P1-P10) exhibit more complex behavior, as shown in Figure S26, Supplementary Movie 1 and 2. A close scrutiny of vibrational analysis highlights two key features: First, the observed mode degeneracies at the $\Gamma$-point lead to cross-dispersion patterns. The symmetry-induced vibrational degeneracies and the associated cross-dispersion patterns suggest possible interactions between the modes with comparable frequencies. Such interactions ensure that these modes are not isolated but instead influenced by nearby vibrational states, which is an important characteristic of coupled phonon modes  \cite{ZHANG2019244}. Second, the variations in amplitudes of different vibrational modes across the layers indicate a non-uniform intensity distribution, implying complex interlayer coupling and vibrational interactions consistent with the experimentally observed mixed nature of the phonon modes.

\begin{figure*}[t!]
	\begin{center}
	\includegraphics[width=0.75\textwidth]{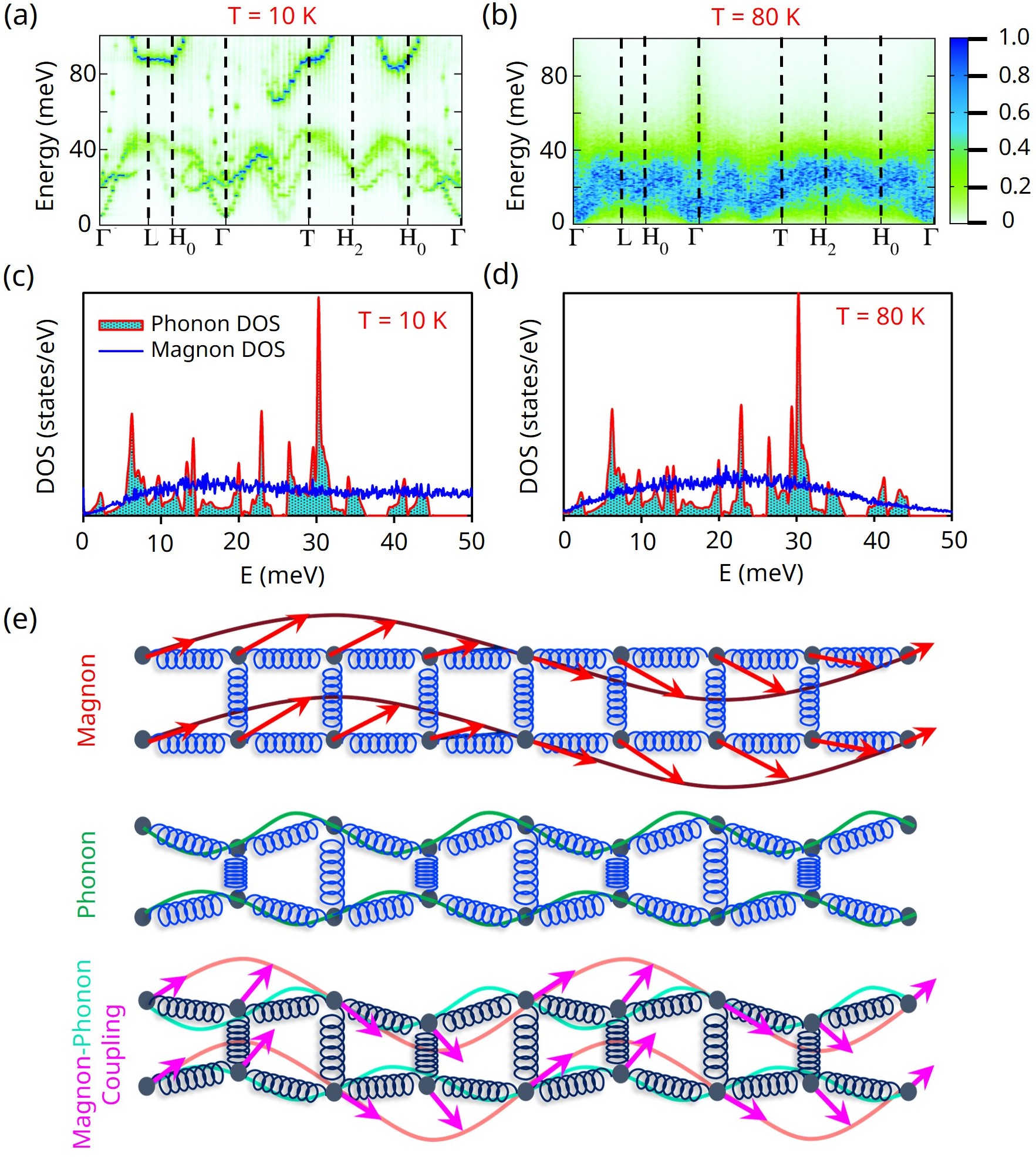}    	
	\end{center}
    \caption{\textbf{Theoretical calculation of magnon and phonon dynamics in \FFGT.} \textbf{(a,b)} Calculated dynamical structure factors, $S(\mathbf{q},\omega)$, plotted along high-symmetry paths for \FFGT at (a) $T = 10$~K and (b) $T = 80$~K. Regions with higher intensity (blue) indicate the adiabatic magnon dispersion lines. \textbf{(c,d)} Calculated phonon and magnon density of states at (c) 10~K and (d) 80~K. \textbf{(e)} Schematic representation of magnon, phonon, and magnon–phonon coupling, illustrating the interaction pathways resolved in the calculations.}
	\label{fig:Fig4}
\end{figure*}

\section*{Temperature-Dependent Raman Spectroscopy in \FFGT} 

Phonon anomalies in Raman spectra, especially at or below the magnetic transitions, are indicative of magnetically driven modifications in lattice dynamics. With increasing temperature, phonon linewidths generally broaden due to enhanced phonon-phonon and electron-phonon scattering~\cite{pp2008, Gran1999, Kle1966, Post1968, Wu2021}. As shown in Fig.~\ref{fig:Fig2} (a,b) and Sections S6.3-S6.5, phonon modes P2–P5 harden continuously upon cooling until  \TSR, followed by anomalous softening down to $\sim$83~K. Below this, no discernible anomaly near $T_Q \sim 50$~K \cite{Pal2024, PalESR2024} is observed, likely due to the limited resolution of the measurement setup and very weak nature of the transition (see Section S5). While phonon linewidths become narrower with decreasing temperature due to reduced phonon-phonon scattering, an anomalous broadening below \TSR suggests additional contributions of magnetic origin, such as spin-phonon coupling. These observations are reproducible and have been verified on flakes of varying thickness (see Sections S7 and S8, including thickness-dependent data in Section S3).

To analyze the phonon anomalies, we consider the temperature dependent change in total phonon frequency as $\Delta \omega(T) = \omega \mathrm(T) - \omega_0 = \Delta \omega_{\mathrm{anh}} + \Delta \omega_{\mathrm{e-ph}} + \Delta \omega_{\mathrm{s\text{-}ph}}$~\cite{Gran1999, Fano1}, where the terms represent contributions from intrinsic phonon-phonon anharmonicity, electron-phonon coupling, and spin-phonon interactions, respectively. The absence of asymmetric Fano line shapes or an electronic background in the phonon modes indicates weak electron-phonon coupling compared to spin-phonon or phonon-phonon interactions below and above~\TSR \cite{Fano1, Fano2}. We therefore isolate the intrinsic anharmonic contribution using the symmetric three-phonon model~\cite{Balkansky1983, Kle1966}. Defining $x = \hbar \omega_0 / (2k_B T)$, the temperature dependence of phonon frequency and linewidth are given by $\omega_{\mathrm{anh}}(T) = \omega_0 - A [1 + 2 / (e^x - 1)]$ and $\Gamma_{\mathrm{anh}}(T) = \Gamma_0 + B [1 + 2 / (e^x - 1)]$, where $\omega_0$ and $\Gamma_0$ are zero-temperature values, and $A$, $B$ are fitting constants. These expressions accurately describe the data between 120--300~K, while deviations below \TSR and above \TC (Fig.~\ref{fig:Fig2}(c,d)), especially for the P4 and P5 modes (Fig.~\ref{fig:Fig3}), suggest a possible presence of spin-phonon coupling at low temperatures and short-range magnetic correlations above \TC, consistent with ESR findings~\cite{PalESR2024}. 

To quantify spin-phonon interactions, we analyze the phonon frequency shifts using a nearest-neighbor approximation in which the renormalization of phonon frequency is proportional to the spin-spin correlation function $\langle \mathbf{S}_i \cdot \mathbf{S}_j \rangle$ and the spin-phonon coupling constant $\lambda_{\mathrm{s\text{-}ph}}$ (see Section S6.7)~\cite{Gran1999, Sus2005, Fen2006, Casto2015, Du2019, Sun2011, Iturriaga2023a}. Since \FFGT\ is a non-collinear ferromagnet \cite{Pal2024, PhysRevB.111.134449}, estimating the spin correlation strength is not straightforward. However, based on the experimental value of the magnetic moment of Fe as $\sim 1.83~\mu_B$/Fe at 83~K~\cite{Pal2024, PalESR2024, Mondal2021}, we estimate $\langle \mathbf{S}_i \cdot \mathbf{S}_j \rangle \approx 0.83$ and $\lambda_{\mathrm{s\text{-}ph}} \approx 0.95~\mathrm{cm}^{-1}$ for the P4 and $0.74~\mathrm{cm}^{-1}$ for P5 (see Table 2), which are substantially higher than those observed in related 2D magnets such as \FthGT\ and Cr$_2$Ge$_2$Te$_6$~\cite{Chen2013, Tian2016, Huang2024}. The large values of $\lambda_{\mathrm{s\text{-}ph}}$ suggest the presence of significant spin-lattice coupling in \FFGT, particularly below \TSR. The decrease in $\lambda_{\mathrm{s\text{-}ph}}$ above \TSR (see Section S6.7 and Table T3) reflects the weakening of spin-phonon correlations with increasing temperature.

Further insight comes from the phonon lifetimes ($\tau$), derived from linewidths using the energy uncertainty relation: $\tau^{-1} = \Gamma(T)/\hbar$~\cite{Kub2000}, where $\Gamma(T)$ is in units of $\mathrm{cm}^{-1}$ and $\hbar = 5.3 \times 10^{-12}~\mathrm{cm}^{-1}\cdot\mathrm{s}$. For P4, the lifetime peaks at $\sim 0.79$~ps near \TSR, decreases to $\sim 0.74$~ps at 83~K, and 0.51~ps at 300~K (see Table 2 and Section S6.6 for other modes). As shown in Fig.~\ref{fig:Fig3}(b,e) (right axes), the inverse relationship between $\lambda_{\mathrm{s\text{-}ph}}$ and $\tau$ confirms that strong spin-phonon coupling shortens phonon lifetimes below \TSR.~\cite{Cho2022, Kim2019, yan2022, mai2021}.

Figure~\ref{fig:Fig3}(c,f) presents the intensity evolution of modes P4 and P5 with temperature, representative of the behavior observed for all low-frequency modes P2–P5 (see Section S6.4 for P2 and P3). The intensity increases steadily up to \TSR, then decreases monotonically until \TC. Above \TC, the intensity resurges, likely driven by short-range magnetic correlations~\cite{PalESR2024,saha-nips3-shortrangecorrelation}. These trends reflect changes in phonon population and the enhancement of spin–phonon coupling near \TC and \TSR~\cite{She1971}. Similar intensity anomalies near \TC\ have been reported in ferromagnetic semiconductors such as CdCr$_2$Se$_4$ and Cr$_2$GeTe$_4$~\cite{Ste1970, Bal1970, Atul2024}, where phonons modulate super-exchange pathways and influence magnetic exchange interactions~\cite{Vaclavkova2020}.

\begin{table*}[ht]
\textbf{Table 2: The fitting parameters for the phonon modes in \FFGT. These parameters were derived using the three-phonon fitting model.}\\
\begin{ruledtabular}
\begin{tabular}{@{}ccccccc@{}} 
\textbf{Mode} & \textbf{$\omega_0$} & \textbf{$A$} & \textbf{$\Gamma_0$} & \textbf{$B$} & \textbf{$\lambda_{s-ph}$} (at 83 K)  & $\tau_{max}$ (at \TSR)\\
& (cm$^{-1}$) & (cm$^{-1}$) & (cm$^{-1}$) & (cm$^{-1}$) & (cm$^{-1}$)  & (pico sec)\\
\midrule

P2    & 100.1 $\pm$ 0.13 & 0.7 $\pm$ 0.03 & 5.6 $\pm$ 0.27 & 0.50 $\pm$ 0.04 & 0.73 & 0.75 \\ 

P3    & 109.2 $\pm$ 0.14 & 0.9 $\pm$ 0.02 & 2.3 $\pm$ 0.09 & 0.30 $\pm$ 0.02 & 1.02 & 1.67 \\ 

P4    & 132.9 $\pm$ 0.11 & 1.1 $\pm$ 0.02 & 4.5 $\pm$ 0.01 & 0.89 $\pm$ 0.001 & 0.95 & 0.79 \\ 

P5    & 148.3 $\pm$ 0.07 & 1.0 $\pm$ 0.02 & 5.8 $\pm$ 0.01 & 0.40 $\pm$ 0.001 & 0.74 & 0.81 \\ 

\end{tabular}
\end{ruledtabular}
\end{table*}

To evaluate the role of spin–phonon (or magnon–phonon) coupling below \TSR, we have analyzed the dynamical spin–spin correlations at 10~K and 80~K. The analysis was carried out through a three-step first-principles-based workflow. First, we determined the magnetic ground state using DFT + DMFT calculations that incorporate spin–orbit coupling (SOC), using temperature-dependent structural inputs from Ref.~\cite{PalESR2024}. The inclusion of dynamic electronic correlations enables a more accurate description of magnetic behavior~\cite{Kagome_PRL} in Fe$_4$GeTe$_2$. The magnetic moments of both symmetric sites of Fe, as presented in Table~T5 of Supporting Information, reveal a strong dependence of spin and orbital magnetic moments on structures. Second, with the converged static magnetic ground state, we have performed a Green’s function-based calculation to extract the intersite exchange interactions in the tensor format after employing a Liechtenstein-Katsnelsen-Antropov-Gubanov (LKAG) formalism combined with the DFT+DMFT methodology \cite{LIECHTENSTEIN198765}. In this method, the description of the low-energy spin excitations were obtained after constructing an effective spin Hamiltonian, which contains terms like symmetric isotropic and anisotropic Heisenberg exchange ($J_{ij}$ and $\Gamma_{ij}$) and the antisymmetric and anisotropic Dzyaloshinskii-Moriya ($D_{ij}$) exchange model Hamiltonian \cite{Kagome_PRB}. Third, with the extracted exchange parameters, a bilinear effective model Hamiltonian is constructed with terms like: 
\begin{equation}
 \mathcal{H}_{\mathrm{mod}}^{i,j} = -J_{ij} \mathbf{s}_i \cdot \mathbf{s}_j - \mathbf{D}_{ij} \cdot (\mathbf{s}_i \times \mathbf{s}_j) - \mathbf{s}_i \cdot \Gamma_{ij} \cdot \mathbf{s}_j - \kappa \sum_{k=i,j} \left( \mathbf{s}_k \cdot \mathbf{s}_k^r \right)^2.
\end{equation}
Here, $\mathbf{s}_i$ and $\mathbf{s}_j$ are the unit vectors along the direction of the spin moments at the atomic sites $i$ and $j$, and $\mathbf{s}_k^r$ is the easy axis, along the arbitrary unit vector $\mathbf{r}$. As per the implementations in the UppASD software, atomistic spin dynamics simulations are performed, where the dynamics of small fluctuations of the spins are studied around the classical local spin moment to converge upon its dynamical magnetic ground state \cite{Etz982458, eriksson2017atomistic}. The presence of low-energy magnons in this system is obtained after Fourier transforming the spin-spin correlation function (the dynamical structure factor) as:
\begin{equation}
S(q,\omega) = \frac{1}{2\pi N} \sum_{i,j} e^{iq(r_i - r_j)} \int_{-\infty}^{\infty} d\tau \, e^{-i\omega \tau} \left\langle \mathbf{s}_i \cdot \mathbf{s}_j(\tau) \right\rangle,
\end{equation}

where, $\mathbf{r}_i$ is the position vector of the magnetic atoms. In Fig.~\ref{fig:Fig4}(a,b), we have plotted the resultant $S(q,\omega)$ values along different high-symmetry directions for two thermal conditions, viz. 10~K and 80~K, without any applied field. The adjacent colour scale presents the normalized values of $S(q,\omega)$, the higher values of which designate the adiabatic magnon dispersion lines. In Fig. S27, the spin component-projected $S(q,\omega)$ plots at temperatures 10~K and 80~K are presented. The energy scales of magnon dispersions reveal their overlap with those of phonon dispersions. This overlap will be more evident from the respective density of states DOS plots for phonon and magnon at both temperatures 10 and 80~K, as depicted in Fig.~\ref{fig:Fig4}(c,d). The magnon DOS plotted in Fig.~\ref{fig:Fig4}(c,d) exhibits a smooth variation with energy in comparison to the phonon ones. For an $N$-atomic system, phonon dispersions have 3$N$ branches per cell, containing many optical modes and the corresponding van Hove singular peaks. In general, magnons contain a lesser number of branches, mostly in the acoustic range, and with a lesser number of van Hove singular peaks. 

For \FFGT, in contrast to the phonon DOS with multiple van Hove maxima from numerous optical branches, the magnon DOS is comparatively featureless, because the spin-wave spectrum as presented in Fig. \ref{fig:Fig4}(a,b) comprises only acoustic-like branches with smooth dispersion in the energy range plotted in both figures. After standard broadening, the resulting magnon DOS varies slowly with energy, which is consistent with the expected $g(\omega) \propto \omega^{1/2}$ behavior at low frequency for typical ferromagnet~\cite{Kittel2004, Blundell}. Thus, the low-energy quasiparticle excitations, like magnons and phonons, access the same energy range, indicating a significant possibility of the occurrence of magnon-phonon coupling in this system. Fig.~\ref{fig:Fig4}(e) presents a schematic illustrating magnon–phonon coupling, whose contribution is crucial for explaining the experimentally observed deviations from the standard anharmonic model.


In conclusion, the temperature-dependent Raman study of the van der Waals ferromagnet \FFGT\ reveals clear signatures of strong spin–phonon coupling and anharmonic lattice dynamics. The polarization-resolved measurements allow us to distinguish a purely out-of-plane Raman mode from several others that exhibit mixed symmetry character, consistent with their complex vibrational origins. First-principles phonon calculations further show that these modes arise from a combination of symmetry-driven degeneracies, interlayer vibrational variations, and cross-dispersion behavior near the $\Gamma$-point, together explaining the mixed nature of the observed phonons. Across the spin-reorientation transition, the phonon modes exhibit anomalous softening, linewidth broadening, and lifetime reduction, consistent with enhanced spin–lattice interaction. At higher temperatures, phonon softening and linewidth broadening reflect dominant phonon–phonon scattering, while deviations from conventional anharmonic behavior persist well above \TC, pointing to robust short-range magnetic correlations. Complementary DFT+DMFT phonon and magnon spectra show overlapping energy scales, providing a natural explanation for the observed magnon–phonon coupling. Together, these results establish \FFGT\ as a model platform for disentangling vibrational and magnetic degrees of freedom in 2D magnetic metals and offer a foundation for tuning spin–lattice interactions for future spintronic and magnonic device applications.

\begin{acknowledgments}

This research has made use of the Thematic Unit of Excellence on Nanodevice Technology (grant no. SR/NM/NS-09/2011) and the Technical Research Centre (TRC) Instrument facilities of S. N. Bose National Centre for Basic Sciences, established under the TRC project of Department of Science and Technology (DST), Govt. of India. D.K. acknowledges the computational resources of BARC supercomputing facility. M.N.H. acknowledges funding from the European Union through the Magnetic Multiscale Modelling Suite (MaMMoS), Project ID: 101135546. D.K. and M.N.H. acknowledge Anwesha Chakraborty for help in the calculation and schematic. A.N.P. acknowledges financial support from DST-Nano Mission Grant No. DST/NM/TUE/QM-10/2019.\\

\end{acknowledgments}

\textbf{Author contributions:}
A.N.P. and R.P. conceived the project. R.P. performed the experiments and carried out the complete data analysis. S.M. and P.M. have grown and characterized the single crystal. Polarization-resolved and initial Raman measurements were done with the help of C.N. under the supervision of A.S. Low temperature (below 77 K) data were collected by M.D. under the supervision of A.M. M.N.H. and D.K. performed the theoretical calculations with help from N.S. and N.P. and wrote the theoretical part. R.P. wrote the initial draft. A.N.P. and P.M. supervised the project, validated the analysis, and reviewed and edited the manuscript. All authors have approved the final version of the manuscript.

\textbf{Competing interests:}
The authors declare no competing interests.

\textbf{Additional information:}
See Supplementary Information for more details.

\textbf{Data availability}: The data that support the findings of this study are available from the corresponding authors upon reasonable request.

\bibliographystyle{naturemag}

\end{document}